# GaN-based Resonant Cavity LEDs Fabricated by Photo-Electrochemical Etching and Micro-Transfer Printing

*Huanqing Chen[1,2], Zhi Li[2,*], Menglai Lei[1], Muhammet Genc[2], Linghai Meng[3], Brendan Roycroft[2], Weihua Chen[1], Xiaodong Hu[1,*], and Brian Corbett[2]*

State Key Laboratory of Artificial Microstructure and Mesoscopic Physics
School of Physics, Peking University
Beijing, 100871 People's Republic of China

Tyndall National Institute, University College Cork
Lee Maltings, Cork, T12 R5CP, Ireland

Guangxi Hurricane Chip Technology Co., Ltd.
Guangxi, 545003 People's Republic of China

***Corresponding Author**, E-mail*:  zhi.li@tyndall.ie and huxd@pku.edu.cn

## Abstract

Resonant cavity LEDs (RCLEDs) exhibit excellent temporal and spatial coherence with narrow spectral linewidth and small divergence angle, which is of great importance for micro-displays. In this paper, we demonstrate a novel method to create GaN-based RCLEDs by using photo-electrochemical etching and micro-transfer 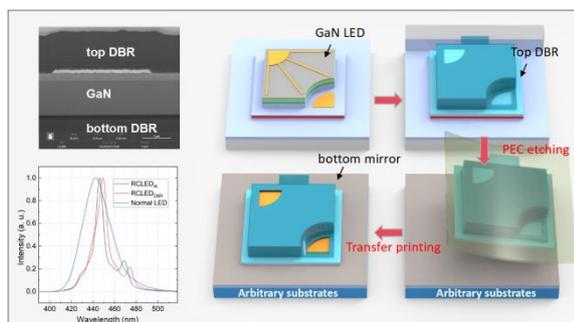 printing (MTP) technology. Through systematic optimizing of the etching conditions, highly selective etching of an InGaN multiple quantum well sacrificial layer is achieved with parasitic etching of adjacent layers being completely suppressed. The roughness of the underside of the exfoliated GaN film is only 3.3 nm. Raman spectroscopy shows that the residual stress in the released material is reduced from 0.74 GPa for the as-grown sample to -0.15 GPa. Using the MTP method, GaN coupons with a deposited upper dielectric mirror were transferred onto target substrates covered with either an Al mirror or dielectric distributed Bragg reflector to form two types of blue RCLEDs. The electroluminescence spectra of the two RCLEDs show a much narrower linewidth, reduced from 32 nm in the conventional LED to ~5 nm, together with stable peak wavelength with increasing current density, with the shift reduced from 9.3 nm to less than 1 nm. The far-field pattern is influenced by the bottom mirror, and the far-field divergence angle can be decreased to only 52° by matching the cavity and the quantum well exciton modes. This scalable approach is highly promising for the realization of compact resonant cavity devices and their use in displays and in communications.

## 1. Introduction

Micro-light-emitting diodes (micro-LEDs) have received wide attention for microdisplays due to their small chip size, high brightness and contrast, and are considered a promising technology for augmented reality and virtual reality (AR/VR)[1,2]. However, there are still some technical problems for micro-LED displays that should be addressed. First, conventional LEDs typically exhibit a standard Lambertian emission pattern with an emission angle of around 120°. When the size is reduced to a micron scale, the proportion of light emitted from the sidewall increases, and the optical crosstalk problem between pixels becomes more serious[3]. For AR/VR applications, where the light needs to be coupled with optical components such as lenses or waveguides, directional light emission (within ±30°) would be preferred[4–6]. The second drawback is related to the LED spectrum, which has a broad linewidth and shifting peak wavelength when the injection current changes[7]. A wide spectrum also results in lower color purity, and the dispersion of LED light diffracted by gratings or lenses could lead to a rainbow phenomenon[8]. In addition, for a large number of micro-LED pixels, the fluctuation of the intrinsic luminescence and driving current of each chip will also bring considerable wavelength shift[9,10].

The spatial and temporal coherence of LEDs can be improved by making a short vertical Fabry-Perot cavity with mirrors on each side to form a resonant-cavity LED (RCLED)[11]. This approach reduces the divergence angle and spectral linewidth of the emitted light, thus achieving directional light emission with higher color purity. Furthermore, wavelength stability is improved since the electroluminescence (EL) wavelength is determined by the cavity mode[12,13]. These advantages make RCLEDs promising light sources for next-generation displays especially in AR/VR applications. One key challenge for GaN-based RCLEDs lies in the mirror fabrication, which is typically achieved using epitaxial growth or by deposition of dielectrics. The first substrate emitting GaN RCLED used an epitaxial distributed Bragg reflector (DBR) of $Al_{0.08}Ga_{0.92}N$/GaN together with a silver top mirror[14]. The AlInN/GaN DBR stack also allows for a highly reflective bottom mirror due to its lattice matching and high refractive index contrast properties, though requiring challenging growth control[15,16]. On the other hand, a dielectric bottom DBRs would solve part of the growth problem but poses a significant process difficulty. Currently, the wafer bonding, substrate removal, and polishing process flow is usually used to expose the atomically flat GaN bottom surface.[17,18] One of the most common substrate removal method is laser lift-off (LLO), on which Zhang et al. has realized green RCLEDs, and even vertical-cavity surface-emitting lasers (VCSELs)[19–21]. Alternatively, electrochemical etching is also an interesting approach to remove the substrate that eliminates the cavity length error caused by grinding and polishing process. Torres et. al. have successfully realized UV-B RCLEDs and optically pumped VCSELs by etching the AlGaN superlattice sacrificial layer[22]. Overall, most of the above RCLED processes involve wafer bonding to support the epitaxial films with thicknesses of a few micrometers. Recently, the micro-transfer printing (MTP) has been widely used in micro-LED processes and is considered a simple and effective solution for mass transfer of microchips[23,24].

In this paper, we realise short cavity RCLEDs by releasing the thin LED structures (with a deposited top dielectric DBR) from the sapphire growth substrate by using photo-electrochemical (PEC) etching, followed by micro-transfer printing the 50×50 μm² devices onto a new substrate coated with either aluminum (Al) or dielectric DBR as the bottom mirror. The resulting two types of RCLEDs showed distinct cavity mode peaks in the electroluminescence (EL) spectrum with wavelengths of 449.3 nm and 445.7 nm and linewidths of 5.6 nm and 5.1 nm, respectively. Thanks to the resonant cavity, the blue shift of the peak wavelength with increasing current densities is significantly reduced from 9.3 nm to less than 1 nm. The evolution of the RCLED far field patterns is studied for different detuning states, and the far-field divergence angle of matched RCLEDs is reduced to only ±26°. To our best knowledge, this is the first demonstration of RCLEDs achieved by micro-transfer printing technology. Compared to the existing techniques, this technology provides a simplified and controlled approach to fabricate RCLEDs and VCSELs.

## 2. Experiment

The cavity element for the GaN-based RCLED was grown on a c-plane sapphire substrate using metal-organic chemical vapor deposition (MOCVD). The epitaxial structure is shown in Fig. 1(a), where following growth of a GaN buffer layer and including a 500-nm-thick n-GaN as the electron transport layer for the PEC reaction, 3 pairs of n-doped InGaN/GaN quantum wells (QWs) were grown as the sacrificial layer (doping concentration = $2\times10^{18}$ cm$^{-3}$). Its thickness was ~20 nm with a photoluminescence (PL) wavelength of ~415 nm designed to partially absorb the 405 nm light from the illumination source during the PEC etching. In order to inhibit the upward diffusion of the photo-generated carriers into the functional layer of the LED, an additional 50 nm of unintentionally doped (UID) GaN was grown on top of the sacrificial layer. The structure of the LED functional layer consists of a Si-doped n-GaN layer, 2 pairs of InGaN/GaN QWs with an emission wavelength of ~450 nm, a 10-nm-thick $Al_{0.25}Ga_{0.75}N$ electron blocking layer, a Mg-doped p-GaN, and a thin layer of heavily-doped p+GaN as a contact layer (doping concentration = $1\times10^{20}$ cm$^{-3}$). The thickness of n-GaN and p-GaN layer is 820 nm and 210 nm, respectively.

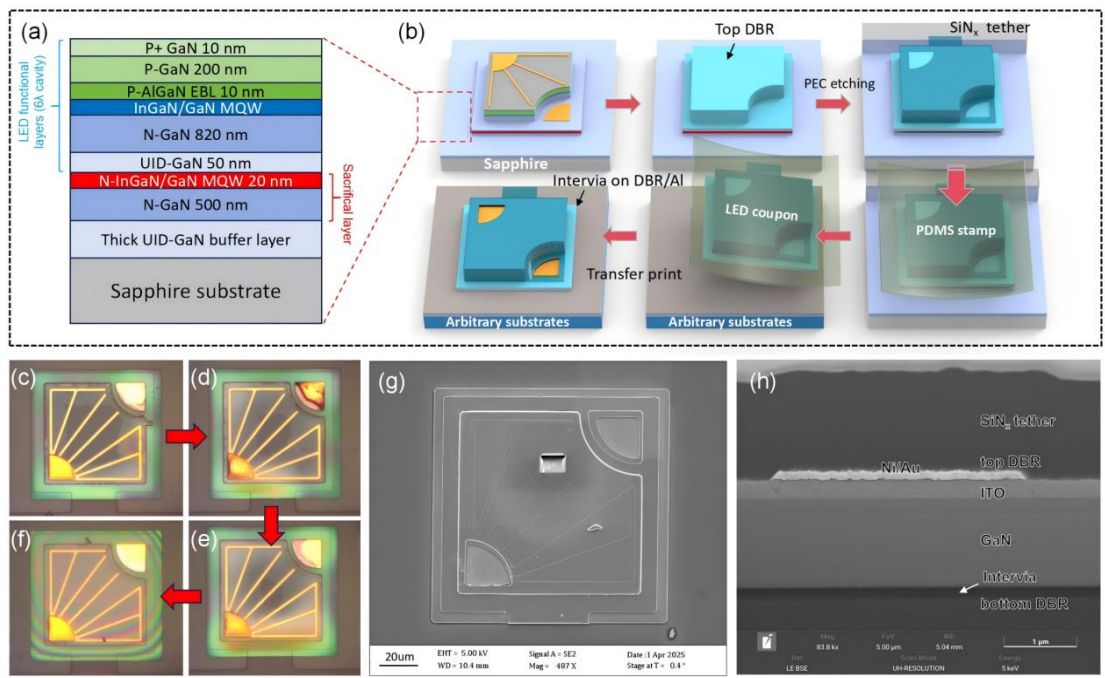

**Figure 1. (a) Schematic of as-grown LED epitaxial layer for PEC etching. (b) The detailed fabrication process flow for the GaN-based RCLED. (c) - (f) Optical microscopy images of LED structure during PEC etching. (g) SEM image of RCLED after transfer printing to new target with DBRs. The hole is due to the FIB cut. (h) Cross-sectional SEM image of the RCLED.**

The schematic of the device process flow is shown in Fig. 1 (b). A standard LED process, comprising of p-contact deposition, mesa etch and n-contact deposition, was first carried out. Then a top DBR coating consisting of 3 pairs of $SiN_x/SiO_2$ (total thickness ~450 nm with nominal peak reflectivity of 60%) was deposited by plasma enhanced chemical vapor deposition (PECVD). The thick DBR layers not only provide ~60% reflectivity for the resonance cavity, but also encapsulate the sidewalls of all the functional layers, especially the QWs, which is important for avoiding their damage during the PEC etching process. To prepare the devices for PEC etching, the top DBR was first patterned, followed by the second mesa etch (slightly wider than the first one) into the bottom n-GaN (~450 nm deep) to expose the sidewalls of the sacrificial MQWs. A 800-nm-thick $SiN_x$ layer was then deposited by PECVD as the anchor/tether material to avoid damage from the corrosion of PEC solution. After PEC etching of the sacrificial MQWs, a 100 nm thick Intervia layer was coated on the new substrates as the adhesive layer. The released LEDs were then picked up by a polydimethylsiloxane (PDMS) stamp and printed on a Silicon target coated with either a nominal 99% reflective DBR (15 pairs of $Ta_2O_5/SiO_2$) or Al film. Finally, two openings above the p and n metal contacts were formed by etching away the DBR layers above them, making the printed RCLED devices ready for testing. Detailed data on DBR and Al films can be found in the Supporting Materials.

Figures 1 (c) - (f) are optical microscope (OM) images of LED-on-sapphire during the PEC process, where the etching front of the sacrificial layer moves from the square boundary to the interior with essentially the same rate in all directions. When the sacrificial layer is completely etched, the OM image [see Figs. 1 (e)] of LED coupons becomes brighter accompanied by Newton's rings from interference. Figure 1 (d) shows the SEM image of the LED chip after being transfer printed to DBR substrate where it can be seen that the LED chip is intact with flat and undamaged surface. A focused ion beam (FIB) cut was used to clarify the cross-section of the RCLED. Figure 1(f) shows the bottom DBR, the 100-nm-thick Intervia, the GaN epitaxial layers, the indium tin oxide (ITO) layer, the Ni/Au electrode, the upper DBR, and the $SiN_x$ tether. It is worth noting that the bonding interface between GaN and DBR is uniform and free of any holes, which proves the integrity and reliability of the RCLED device.

### 3. Result and Discussion

### 3.1 PEC etching of GaN on sapphire

A major advantage of PEC etching is its selectivity to the energy band, with the photo-generated carriers generated in the sacrificial layer resulting in the etching reaction[25]. However, the etching conditions in practical experiments still need to be finely chosen to ensure a sufficiently large etch selectivity ratio and avoid parasitic etching effects in the GaN layers. Here the irradiation density of an external LED source with emission centered at 405 nm is fixed to 105 mW cm$^{-2}$, a 0.5 mol/L $HNO_3$ solution was used as the electrolyte while different bias voltages were applied on the GaN anode.

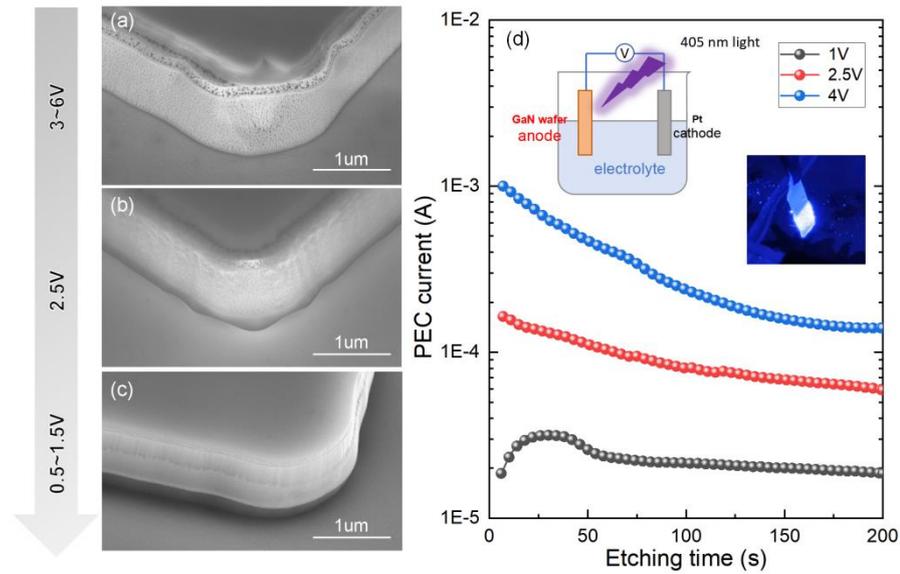

**Figure 2. (a) Typical SEM images of the n-GaN sidewall parasitically etched under applied voltages of 3-6 V, (b) 2.5 V, and (c) 0.5-1 V. (d) Photocurrent versus. time for samples etched under different applied bias voltages. The insets show a schematic of the experiment and a photo during the PEC etching.**

It can be seen in Fig. 2 (a) that when the applied bias voltage is in the range of 3-6 V, not only is the sacrificial layer etched, but also the sidewalls of the n-GaN layer above it show a large number of pores (20~30 nm in diameter), which are believed to extend to the inside of the n-GaN[26]. Although some literature suggests that porous GaN is effective in modulating the refractive index of the material to achieve optically confined or highly reflective DBRs, the severe scattering effect due to the porous structure makes it very unfavorable for the RCLED devices in this work[27,28]. The parasitic etching is due to the excessive tilting of the n-type GaN energy bands by the applied bias, resulting in an increase in the hole concentration by tunneling effect[29,30]. Therefore, a lower applied bias is required to eliminate this effect. As shown in Fig. 2 (b), when the voltage is reduced to 2.5 V, the diameter of pores in the sidewall become significantly smaller. Furthermore, when the voltage is lowered to 1 V, the sidewalls are smooth as originally etched with no pores. Figure 2 (d) shows the photocurrent versus time profile throughout the PEC etching process. It can be seen that for both 4 V and 2.5 V the photocurrent gradually decays exponentially as the reaction proceeds, and a non-zero photocurrent still exists after the etching is completed. On the other hand, for an applied bias voltage of 1 V the photocurrent first increases and then decays during the undercutting. We believe that when a small bias voltage is applied, the reaction needs to overcome some initial barrier such as the oxide layer on the surface and thus the current gradually increases. Then when the sacrificial layer is undercut further, the conducting cross-sectional area becomes smaller and the current decays[30]. The etching rate for the three voltages was approx. 16 μm/min, 11 μm/min, and 3 μm/min, respectively. We also reduced the concentration of the electrolyte, but it had little effect on inhibiting the formation of pores in n-GaN (not shown here). Therefore, the optimum PEC etching voltage and electrolyte concentration were determined to be 1 V and 0.5 mol/L, respectively.

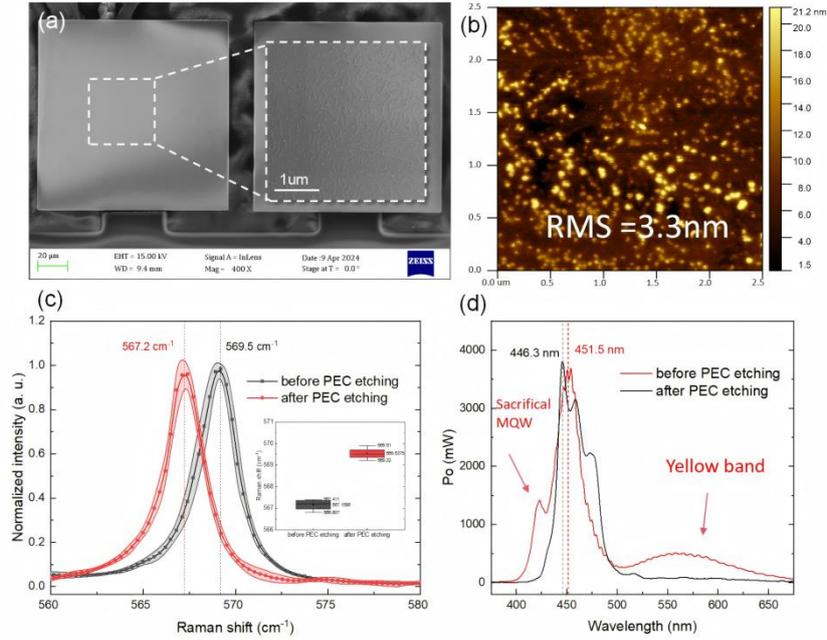

**Figure 3.** (a) SEM image of the backside of the released LED coupon. (b) AFM image of the released N-face GaN. (c) Raman spectra and (d) PL spectra of the samples before and after fully PEC etching. The inset is a box plot of the $E_2$ (high) mode peak.

The quality factor of the resonant cavity is greatly affected by the scattering loss due to the roughness of the interfaces. To characterize the morphology of the LED backside, the LED coupon was flipped using a PDMS stamp for evaluation by secondary electron microscopy (SEM) and atomic force microscopy (AFM). It can be clearly seen in Fig. 3 (a) that the surface after PEC etching is flat and smooth on a large scale. The higher magnification inset shows the presence of a mesh structure on the backside, which might originates from the current path during the PEC etching process. The AFM image in Fig. 3 (b) shows that the backside mesh structure consists of clustered small particles with diameters of 50~100 nm. These particles are likely to be a residual intermediate product such as $Ga_2O_3$[31]. Nevertheless, the root mean square (RMS) roughness of the backside is still as low as 3.3 nm, and there is no parasitic etching effect on the backside of n-GaN. Compared to wet etching method for GaN-on-Si or LLO method for GaN-on-Sapphire to remove the substrate, a much smoother surface can be obtained using PEC etching because of the better quality and the sharper interface of the grown InGaN MQW sacrificial layer. We believe that the RMS roughness of the etched surface can be improved further by optimizing the PEC etch conditions and/or performing post-treatment[32]

For heterogeneous epitaxy of GaN, heavy internal stresses from lattice mismatch and thermal mismatch are a major problem. As a typical example, GaN grown on sapphire substrate is usually subjected to compressive stresses, which lead to strong piezoelectric polarization and severe quantum confined Stark effect (QCSE) effects on the output light[33]. To verify the effect of stress release, micro-Raman and photoluminescence (PL) was performed on the same LED chip before and after PEC etching. As shown in Fig. 3 (c), the average $E_2$ (high) mode frequency moves from ~569.5 cm$^{-1}$ (before lift-off) to ~567.2 cm$^{-1}$ after complete release, which is very close to the stress-free state of bulk GaN[34]. This result is confirmed by measuring a large number of samples. The residual stresses in the LED epitaxial layer before and after releasing were calculated as 0.74 GPa and -0.15 GPa, respectively. The results show that the PEC etching method has a significant effect on the compressive stress release in the GaN layer due to the complete elimination of the connection and lattice mismatch between the substrate and the epitaxial layer. Meanwhile, the PL spectra of two samples also imply similar conclusions. As shown in Fig. 3 (d), after the etching of the sacrificial layer, the spectrum of the MQWs shows multiple Fabry-Perot peaks. The wavelength of the shortest peak is 446.3 nm, showing a blue shift of ~5 nm compared to the non-released material. The luminescence peaks associated with the sacrificial layer around 420 nm and the GaN yellow band emission at ~550 nm also disappear, indicating that the substrate and the GaN buffer layer were completely removed. The blue shift of the emitting peaks might be caused by the reduction of the stress-induced energy band tilting effect after removing the substrate[35,36].

**3.2 The optoelectronic performance of RCLEDs**

LEDs with top DBRs were then transferred onto two high reflective targets covered with either an Al film or DBR coating to form $RCLED_{Al}$ and $RCLED_{DBR}$ respectively. The optoelectronic properties of normal LEDs (no transfer), $RCLED_{DBR}$, and $RCLED_{Al}$ were performed under room temperature (RT) and continuous wave (CW) operation. Figure 4 (a) shows the I-V curves of the three LEDs. It can be seen that the forward voltage of $RCLED_{DBR}$ at the injected current of 20 mA is reduced by 0.18 V compared to $RCLED_{Al}$ (from 4.50 V to 4.32 V), which is mainly brought by the reduction in series resistance. Since the thermal conductivity of DBR composed of dielectric material is much lower than that of Al metal, a large amount of heat will accumulate in the $RCLED_{DBR}$, resulting in higher temperatures and lower voltage in the device. This result is also evidenced by the power rollover of the $RCLED_{DBR}$ at fairly high currents ~80 mA [see the inset of Fig. 4 (b)]. On the other hand, the voltage of control LED is ~4.40 V at 20 mA, which is slightly lower than that of $RCLED_{Al}$. As shown in the inset of Fig 4 (a), the reverse leakage of both RCLEDs is greatly reduced from 3 μA to around 0.6 μA at -5 V. A possible reason of the decrease in the forward voltage of the control LED is the increase of the leakage channel[37]. Since the PEC etching cuts all the connection between the LED and the substrate, effectively reducing the number of leakage channels to the buffer GaN layer with high dislocation density.

The light output power (LOP) of the three samples as a function of injected current are shown in Fig. 4 (b). The two RCLEDs exhibit almost the same output power at 20 mA, however, their light output power is 0.6 times smaller than that of normal LEDs. However, there is better spectral coherence in the RCLEDs[38]. The inclusion of a resonant cavity tighten the selection conditions for photons emitted from the active region, only photons near the resonant mode are allowed to exit outwards. Although the integral LOP of both RCLEDs is smaller compared to control LEDs, the EL images [see Fig. 4(c) and (d)] show that the RCLEDs emit a much brighter and more concentrated light from the topside. This confirms that RCLEDs with cavity effect performs a smaller far-field divergence angle and better directionality of light emission, which is very beneficial in applications like AR/VR display and visible light communication (VLC).

Another signature feature of RCLEDs associated with the Fabry–Pérot cavity is the width narrowing of emission spectrum. Figure 4 (g) shows the EL spectra of the three LEDs at different injection currents, it can be seen that the normal LED has a broad peak of spontaneous emission. A blue shift of the peak wavelength as well as the peak broadening with increasing injection current can be clearly observed. The total peak blue shift of is ~5.8 nm from 448.9 nm to 443.1 nm with full width at half maxima (FWHM) increases from 32 nm to 39 nm when the injection current is increased from 2 mA to 15 mA. A well-recognized reason for the blueshift of MQWs as current increasing is considered to be the screening effect of injected carriers on the polarization electric field, which compensates for the QCSE effect and flattens the energy band[39]. The wavelength shift of LEDs can bring about color deviation and unevenness, which is very detrimental to the display. In contrast, multiple sharp and stable mode peaks are clearly observed in both RCLEDs, the main peak FWHM of $RCLED_{DBR}$ and $RCLED_{Al}$ are reduced to 5.1 nm and 5.6 nm, respectively, demonstrating the wavelength selection effect of the resonant cavity structure. Meanwhile, the main peak wavelength of the $RCLED_{DBR}$ is near 449.3 nm with a mode spacing of about 22.9 nm, whereas the main peak wavelength of the $RCLED_{Al}$ is shorter at 445.7 nm and the mode spacing is smaller at 24.2 nm. The difference in mode spacing of the two RCLEDs is most likely due to the greater depth of penetration of the DBR as compared to that of the Al. In addition, it can be seen from Fig. 4 (h) that the EL spectra of RCLEDs have strong current stability. The variation of half-width and wavelength of the main peak remain almost within 1 nm when the current is increased from 5 mA to 15 mA. The presence of the resonant cavity not only allows LEDs to have higher color purity, but also substantially relaxes manufacturing tolerances brought about by uniformity of epitaxy and working environment.

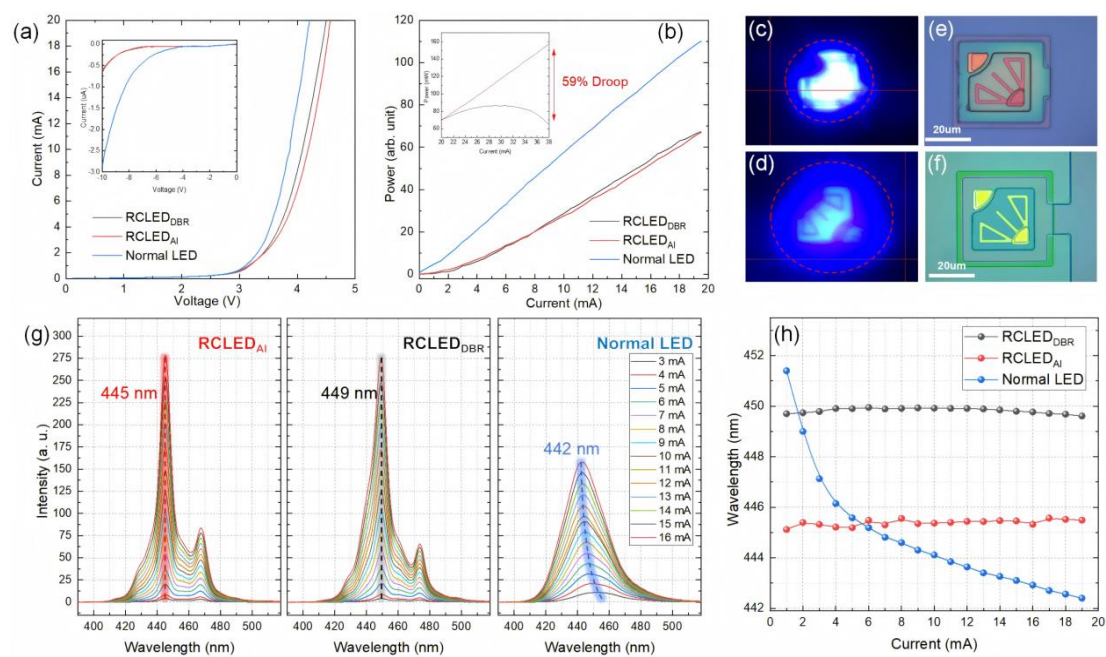

**Figure 4.** (a) Forward I-V and reverse I-V (inset) characteristics curves of RCLEDs and normal LED. (b) L-I characteristics of RCLEDs and normal LED. (c) Near-field EL emission images of RCLED and (d) normal LED. (e) OM images of RCLED and (f) normal LED. (g) EL emission spectra of RCLEDs and normal LED. (h) EL peak shift of RCLEDs and normal LED as a function of injection current.

Figures 5 (a) and (b) show the angle-resolved PL (ARPL) spectra of $RCLED_{DBR}$ and $RCLED_{Al}$ under 355 nm laser pumping. Parabolic dispersion curves are clearly observed in both RCLEDs, with the peaks of the resonant modes shifting to higher energies as the angle increases, which indicates the coupling of the MQW emission into the resonant cavity mode[40]. The two RCLED samples show different far-field distributions. For $RCLED_{DBR}$, the intensity maximum in ARPL is located at both sides of the parabola, whereas in the RCLED sample with the Al mirror, the intensity maxima appear at the bottom of the parabola, i.e., at θ = 0°. This is due to the fact that aluminum and dielectric DBR have different phase conditions for reflection[41]. As shown in Fig. 5 (c), the Al mirror interface in the resonant cavity is always located at the antinode of the optical field. However, for the RCLED$_{DBR}$, the antinode of the wave intensity is located at the interface from SiO$_2$ to Ta$_2$O$_5$. Therefore, the consequent phase difference results in a shift of the resonant wavelength of RCLED$_{DBR}$ to a longer region. Fig. 5 (b) shows the far-field divergence angle fitted to the ARPL data. Due to the cavity mode matching effect, the far-field radiation pattern of RCLED$_{DBR}$ shows an emission peak at an angle of ±25.3° to the normal, with a slight depression of the intensity at the center, forming a heart shape[42,43]. While the far-field pattern of RCLED$_{Al}$ shows a maximum emission in the normal direction, and the divergence angle is drastically decreased to about 52°, which is significantly smaller than the half-angle of 120° from the conventional Lambertian light sources. Further reduction in divergent angle can be obtained by increasing the pairs of the top DBR or increasing the cavity length[44–46].

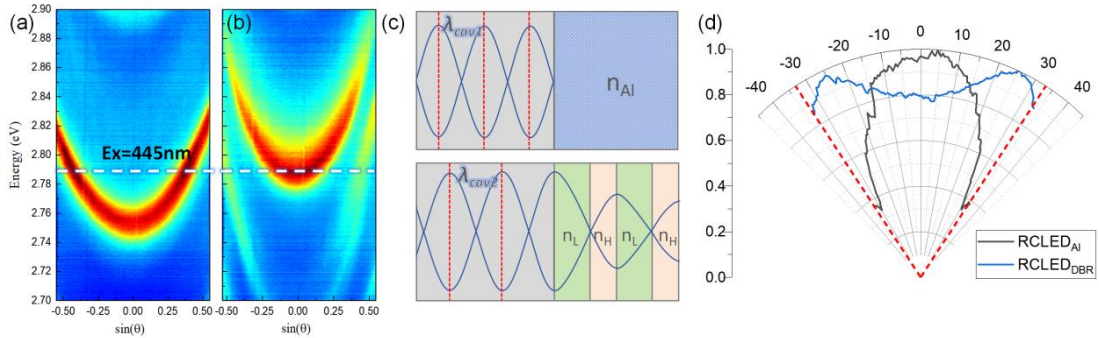

Figure 5. (a) Angle-resolved PL spectra of RCLED$_{DBR}$, and (b) RCLED$_{Al}$. (c) Electric field phase distribution at the interface in the resonant cavity with Al and BDR mirror. (d) Far-field emission patterns of RCLED$_{DBR}$ and RCLED$_{Al}$.

## 4. Conclusion

In summary, we have proposed a new approach for the fabrication of GaN-based RCLEDs by PEC etching and MTP technology. The devices leverage the high selectivity of PEC etching, which greatly reduces GaN surface roughness and effectively address the challenge of controlling cavity length with traditional RCLED fabrication approaches. Meanwhile, the MTP process not only allows for more flexible integration of the lower DBR, but also provides finer control of device arrays. Two types of RCLEDs with the size of 50×50 μm$^2$ were prepared with different bottom mirrors, and their optoelectronic properties were investigated in detail. The EL spectra of RCLED$_{Al}$ and RCLED$_{DBR}$ showed significant mode peaks at 445.7 nm and 449.3 nm, with linewidths reduced to 5.6 nm and 5.1 nm. The blue shift of the peak wavelength is significantly reduced from 9.3 nm for normal LED to less than 1 nm for both RCLEDs. Moreover, the reflection phase difference between Al and DBR mirrors leads to cavity mismatch phenomenon. RCLED$_{Al}$ exhibits a highly directional emission with the divergence angle of only 52°, while RCLED$_{DBR}$ shows a heart shape in far field pattern due to the negative detuning. The achievements have significantly reduced the difficulty and complexity of RCLED preparation, and we believe that this technology will provide a stable solution for large-scale production of GaN-based RCLEDs and VCSELs.

**Supporting Information**

The detailed structure of RCLED's top DBR and bottom DBR coatings, the reflectance spectra of top DBR and bottom DBR/Al mirrors for RCLEDs, the optical microscope images of substrate transfer process by PEC etching and MTP, as well as the demonstration of RCLED arrays by MTP.

**Funding Sources**

This work was supported by the National Key Research and Development Program of China (Grant No. 2023YFB4604400), Research Ireland Pathway Program (SFI-IRC_22/PATH-S/10800) and Irish Photonics Integration Centre (IPIC) (SFI-12/RC/2276_P2_IPIC).

**Notes**

The authors declare no competing financial interest.